\documentclass[pra,aps,floats,letterpaper,floatfix,footinbib,preprintnumbers,twocolumn]{revtex4}
\usepackage{amsmath}
\usepackage{hyperref}
\usepackage{color}
\usepackage[mathenv]{}
\usepackage{graphicx}
\usepackage{natbib}
\begin{document}

\title{Frequency-Dependent Squeeze Amplitude Attenuation and Squeeze Angle Rotation by Electromagnetically Induced Transparency for Gravitational Wave Interferometers}

\author{Eugeniy E. Mikhailov}
\author{Keisuke Goda}
\author{Thomas Corbitt}
\author{Nergis Mavalvala}

\affiliation{LIGO Laboratory, Massachusetts Institute of Technology, Cambridge, MA 02139, USA}

\begin{abstract}
We study the effects of frequency-dependent squeeze amplitude
attenuation and squeeze angle rotation by electromagnetically
induced transparency (EIT) on gravitational wave (GW)
interferometers. We propose the use of low-pass, band-pass, and
high-pass EIT filters, an S-shaped EIT filter, and an intra-cavity
EIT filter to generate frequency-dependent squeezing for injection
into the antisymmetric port of GW interferometers. We find that
the EIT filters have several advantages over the previous filter
designs with regard to optical losses, compactness, and the
tunability of the filter linewidth.
\end{abstract}

\maketitle

\section{Introdcuction}
The gravitational wave (GW) community is currently exploring
strategies to overcome the quantum limited sensitivity of next
generation GW interferometers. One of the most promising
techniques is the injection of squeezed fields into the
antisymmetric (dark) port of the GW
interferometers~\cite{caves1981prd}. Future GW interferometers
will use higher circulating power to reduce shot noise at high
frequencies, but this increase in power makes the radiation
pressure noise significant at low frequencies. The effect of the
radiation pressure noise is to ponderomotively squeeze the optical
fields with a frequency dependent (FD) squeeze angle due to the FD
response of the test masses~\cite{kimble2002prd}. This presents
difficulty in injecting squeezed states into the interferometer
because the squeeze angle of the squeezed state source must be
matched to the ponderomotive squeeze angle. Squeezed state sources
are generally frequency independent, but the desired angle may be
produced by using optical cavities placed between the squeezed
state source and the interferometer as filters. Both filters that
rotate the squeeze angle to match the ponderomotive squeeze
angle~\cite{kimble2002prd,harms2004prd} and filters that attenuate
the anti-squeezing in a desired band~\cite{corbitt2004prd} have
been proposed. The attenuation filters operate on the principle
that the squeezed state with a frequency independent squeeze angle
has beneficial effects over some frequency band, but harmful
effects over other frequency bands, and those harmful effects can
be mitigated by introducing a FD optical loss. The previously
proposed cavity filters~\cite{kimble2002prd,corbitt2004prd}, while
producing the desired effects, are difficult to build because they
require cavities with narrow linewidths comparable to the
bandwidth of the interferometer ($100~\rm{Hz}$), which require
either kilometer-scale cavities, or high finesse cavities. For
squeeze angle rotation, high finesse cavities are somewhat
undesirable due to high optical losses. For example, assuming a
round trip optical loss of $20\times 10^{-6}$, a cavity would have
to be nearly $100~ \rm{m}$ to maintain an effective loss of $10
\%$ or less. In this paper, we propose alternative filter designs
based on electromagnetically induced transparency (EIT) media, for
both attenuation and rotation, that do not require long or high
finesse cavities, and also have increased tunability.

Several properties of EIT media make them a potentially
advantageous alternative to ultralow-loss or km-scale optical
cavities. First, EIT media typically have narrow transmission
resonance linewidths; the narrowest linewidth reported is $1.3$~Hz
in a paraffin coated cell with $^{85}$Rb as the EIT
medium~\cite{budker99}. The EIT media can also be very compact,
with typical lengths of $\sim 10$~cm. Second, the resonance
linewidth may be adjusted in non-invasive ways, by changing the
atomic density~\cite{lukin97prl} or drive field
intensity~\cite{harris'97pt}, to optimize the filtering. Altering
the linewidth of an optical cavity, on the other hand, usually
requires changing the transmission of the mirrors or the
macroscopic cavity length. Other variable reflectivity techniques,
e.g., three-mirror cavities, are possible, but issues of fringe
control, scattered light and mode mismatch are compounded in such
systems. Third, EIT media can be nearly 100\% transmissive for the
probe field~\cite{xia1999pra,braje2003pra}. Fourth, mode matching
of the spatial mode of the squeezed (probe) beam to that of the
EIT is not critical, as it is for an optical cavity, where mode
mismatch can be a significant source of optical
loss~\footnote{See, Fig. 5 of Ref.~\cite{xia1999pra}, e.g., where
the authors showed that the probe field (in our case a squeezed
state) need not be modematched to drive laser spatial mode, but be
just contained within this mode.}. Fifth, EIT media have also been
shown to preserve squeezed states in
transmission~\cite{akamatsu2004prl}. These reasons lead naturally
to consideration of using EIT for squeezed state filters.

\begin{figure}
\includegraphics[width=0.8\columnwidth]
    {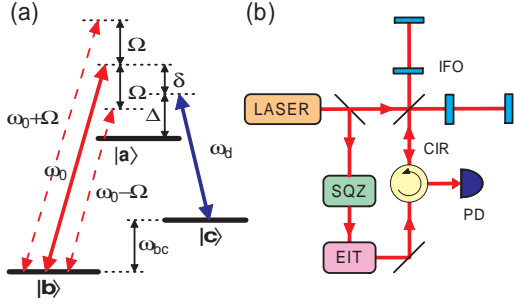}
\caption{
    \label{fig:fig1}
    (a) A generic three-level EIT system. The susceptibility of the EIT
    medium, $\chi$, is a function of the frequency of the probe field,
    $\omega_0$,
    the sideband frequency, $\Omega$ (relative to the carrier frequency of the
    probe, $\omega_0$), the frequency of ground state splitting, $\omega_{bc}$,
    the two-photon detuning of the probe field, $\delta=\omega_0 -\omega_d-\omega_{bc}$,
    the one-photon detuning of the drive field, $\Delta$, and the length of
    the EIT medium, $l$.
    (b) A proposed configuration for a squeeze EIT filter and GW
    interferometer, where SQZ: squeezed field generator; EIT:
    EIT filter; IFO: GW interferometer; CIR: circulator;
    PD: photodetector.
    }
\end{figure}

The paper is organized as follows: In Section~\ref{sect:theory} we
develop the theoretical formulation of how an EIT medium gives
rise to squeeze state attenuation and squeeze angle rotation, and
we derive the quadrature noise at the output of an example
gravitational wave interferometer for a generic input squeezed
state; in Section~\ref{sect:eit} we analyze three different cases
of EIT filtering of the input squeezed state and derive the
quantum noise limited sensitivity of the GW interferometer for
each case; we summarize our findings in
Section~\ref{sect:conclusions}.

\section{Theory}
\label{sect:theory}

In this Section we relate the general properties of the EIT
medium, as described by a complex transmission coefficient, to the
propagation of the squeezed state, in terms of rotation and
attenuation. We then apply this formulation to the output of a GW
interferometer.

\subsection{Squeeze amplitude attenuation and squeeze angle rotation by EIT}

The energy level scheme for one photon resonant EIT is shown in
Figure~\ref{fig:fig1}(a). Using the parameters defined in
Figure~\ref{fig:fig1}, the EIT transmission as a function of the
sideband frequency, $\Omega$, is given by
\begin{eqnarray}
\label{eq:transparency_p} {\mathcal T}(\omega_0+\Omega) = {\rm
exp}\left(i\,\int^l_0 k_{{\rm eit}}(\omega_0+\Omega,z)\,dz\right),
\end{eqnarray}
where $k_{{\rm eit}}$ is the wave vector of the probe field
through the EIT medium. For most cases, it is reasonable to assume
that the susceptibility of the medium $|\chi(\omega_0+\Omega,z)|
\ll 1$ and $\Omega \ll \omega_0$, then the wave vector can be
written as
\begin{eqnarray}
\label{eq:wavevector}
k_{{\rm eit}}(\omega_0+\Omega,z) &=& \frac{\omega_0+\Omega}{c}\sqrt{1+\chi(\omega_0+\Omega,z)} \nonumber\\
&\simeq& \frac{\omega_0}{c}\left(1+\frac{\chi(\Omega,z)}{2}\right).
\end{eqnarray}
In the frequency domain, substituting Eq.~\eqref{eq:wavevector}
into Eq.~\eqref{eq:transparency_p} and moving into the rotating
frame of the probe field, $a$, with frequency,
$\omega_0$~\cite{collett1984pra}, we find the EIT transmission at
sideband frequencies, $\pm\Omega$,
\begin{eqnarray}
\label{eq:transmission_rotating} {\mathcal T}(\pm \Omega) =
{\mathcal T}_{\pm} {\rm e}^{i\, \Theta_{\pm}},
\end{eqnarray}
where
\begin{eqnarray}
\label{eq:trans_chi1}
{\mathcal T}_{\pm} &\equiv& {\rm exp}\left(-\frac{\omega_0}{2c}\,\int_0^l \chi_2(\pm\Omega,z)\,dz\right),\\
\label{eq:theta_chi2}
\Theta_{\pm} &\equiv&
\frac{\omega_0}{2c}\int_0^l\chi_1(\pm\Omega,z)dz.
\end{eqnarray}
Here $\chi_1(\Omega,z)$ and $\chi_2(\Omega,z)$ are the real and
imaginary parts of $\chi(\Omega,z)$. $\chi_1(\Omega,z)$ is
responsible for phase shift and $\chi_2(\Omega,z)$ is responsible
for signal strength attenuation.

When the input field, $a$, is injected into an EIT medium, the
output field, $b$, and its adjoint, $b^{\dagger}$, are given in
terms of $a$ and its adjoint, $a^{\dagger}$, by
\begin{eqnarray}
\label{eq:output}
\hspace{-0.5cm}
b(\Omega) &=& {\mathcal T}(\Omega)\,a(\Omega) + {\mathcal L}_ + (\Omega)\, v(\Omega), \\
\hspace{-0.5cm} b^{\dagger}(-\Omega) &=& {\mathcal
T}^*(-\Omega)\,a^{\dagger}(-\Omega) + {\mathcal
L}^*_{-}(-\Omega)\, v^{\dagger}(-\Omega),
\end{eqnarray}
where ${\mathcal L}_\pm$ is the EIT absorption coefficient such
that
\begin{eqnarray}
\label{eq:loss}
{\mathcal L}_\pm = \sqrt{1 - {\mathcal T}_\pm^2}
\end{eqnarray}
and $v$ is
a vacuum field coupled in by the absorption loss. $a(\Omega)$ and
$a^{\dagger}(-\Omega)$ satisfy the commutation relations
\begin{equation}
\label{eq:commutation_relations} \left[a(\pm \Omega),
a^{\dagger}(\pm \Omega')\right] = 2\pi\delta(\Omega-\Omega^{'}),
\end{equation}
and all others vanish (similarly for $v(\Omega)$ and
$v^{\dagger}(-\Omega)$). In the two-photon representation, the
amplitude and phase quadratures of $a$ are defined as
\begin{eqnarray}
\label{eq:quadratre_fields}
a_1(\Omega) &=& \frac{a(\Omega) + a^{\dagger}(-\Omega)}{\sqrt{2}}, \\
a_2(\Omega) &=& \frac{a(\Omega) - a^{\dagger}(-\Omega)}{\sqrt{2}i},
\end{eqnarray}
respectively (similarly for $b$ and $v$). We find the amplitude
and phase quadrature fields of the output, in compact matrix form,
to be
\begin{eqnarray}
\label{eq:a}
\textbf{b} = \textbf{M}\textbf{a} + \sqrt{1-\left({\mathcal A}_{+}^{2}+{\mathcal A}_{-}^{2}\right)}\textbf{v}_s,
\end{eqnarray}
where we use the two-photon matrix representation
\begin{eqnarray}\hspace{-0.5cm}
\textbf{a} &\equiv& \left(\begin{array}{ccc}
a_1 \\
a_2
\end{array}\right)
\end{eqnarray}
for the operator, $a$ (and similarly for $b$ and $v$), and
\begin{eqnarray}
\textbf{M} = {\rm e}^{i\,\varphi_{-}}\left(\begin{array}{ccc}
\cos\varphi_{+} & -\sin\varphi_{+} \\
\sin\varphi_{+} & \cos\varphi_{+}
\end{array}\right)
\left(\begin{array}{ccc}
{\mathcal A}_{+} & i\,{\mathcal A}_{-} \\
-i\,{\mathcal A}_{-} & {\mathcal A}_{+}
\end{array}\right)
\end{eqnarray}
is a matrix representing propagation through the EIT medium.
$\textbf{M}$ comprises an overall phase shift, $\varphi_{-}$,
rotation by angle, $\varphi_{+}$, and attenuation by a factor,
${\mathcal A}_{+}$. Here we have defined
\begin{eqnarray}
\varphi_{\pm} \equiv \frac{1}{2}\left(\Theta_+ \pm \Theta_-\right),
\ \ \  {\mathcal A}_{\pm} \equiv \frac{1}{2}\left({\mathcal T}_+\pm
{\mathcal T}_-\right),
\end{eqnarray}
and performed a unitary transformation on $\textbf{v}$, such that
\begin{eqnarray}
\sqrt{1-\left({\mathcal A}_{+}^{2}+{\mathcal A}_{-}^{2}\right)}\,\textbf{v}_s &=& \nonumber\\
&&\hspace{-3cm}\frac{1}{2}
\left(\begin{array}{ccc}
{\mathcal L}_+ + {\mathcal L}_- & i\left( {\mathcal L}_+ - {\mathcal L}_-\right)\\
-i\left({\mathcal L}_+ - {\mathcal L}_-\right)& {\mathcal L}_+ +
{\mathcal L}_-
\end{array}\right)\textbf{v}
\end{eqnarray}
and $\textbf{v}_s$ behaves as ordinary unsqueezed vacuum. For
symmetrical lineshapes with respect to the carrier,
$\chi_1(\Omega) = -\chi_1(-\Omega)$, and $\varphi_{+}$ therefore
vanishes, giving no quadrature angle rotation, but attenuating the
signal strength. For asymmetrical lineshapes, nonzero
$\varphi_{+}$ gives quadrature angle rotation.

\subsection{Application to GW interferometers}
\label{sect:GWI}

For a conventional GW interferometer with arm lengths $L$ and
mirror masses $m$, the Fourier transform $h_{n}=h_{n}(\Omega)$ of
the optical noise in the GW strain signal when a (EIT-filtered)
squeezed field $\textbf{b}$ is injected into the dark port is
given by~\cite{kimble2002prd}
\begin{eqnarray}
\label{eq:h} h_{n}(\Omega) = \sqrt{\frac{4\,\hbar}{m\Omega^2 L^2
\cal K}}(b_2 - {\cal K} b_1)\,{\rm e}^{i\tan^{-1}(\Omega/\gamma)},
\end{eqnarray}
and
\begin{eqnarray}
{\cal K}(\Omega) =
\frac{8\,I_0\,\omega_0}{m\,L^2\,\Omega^2(\gamma^2+\Omega^2)}
\end{eqnarray}
is the effective coupling constant that relates motion of the
mirrors to the output signal. Here $\gamma$ is the linewidth of
the arm cavities (typically $\gamma \sim 2\pi\times 100$ Hz),
$\omega_0$ is the carrier frequency of the incident laser light,
and $I_0$ is the laser power at the beam-splitter.

Assuming no losses other than those associated with the EIT
filter, a squeezed state with squeeze factor $r$ and squeeze angle
$\theta = \theta_0 + \varphi_{+}$ is injected into the
antisymmetric port of the GW interferometer through the EIT filter
[see Figure~\ref{fig:fig1}(b)]. The input angle $\theta_0$ may be
arbitrarily chosen by microscopic variations in the distance
between the squeezed state source and the interferometer. The
spectral density of the noise at the output of the GW
interferometer is then
\begin{eqnarray}
\label{eq:S} S_{h} = \frac{4\,\hbar}{m \Omega^2 L^2}\left({\cal
K}+\frac{1}{{\cal K}}\right)V_{\theta+\Phi},
\end{eqnarray}
where
$ V_{\theta+\Phi} = V_{+}\, \sin^{2}\left(\theta+\Phi\right) +
V_{-}\, \cos^{2}\left(\theta+\Phi\right)$.
Here
\begin{eqnarray}
\Phi(\Omega) = \cot^{-1}{\cal K}(\Omega)
\end{eqnarray}
is the effective ponderomotive squeeze angle of the
interferometer, and the noise in the anti-squeezed and squeezed
quadratures, $V_+$ and $V_-$, respectively, is given by
\begin{eqnarray}
\left(\begin{array}{ccc}
V_+ \\
V_-
\end{array}\right)
&=&
\left(\begin{array}{ccc}
{\mathcal A}_+^2&{\mathcal A}_-^2\\
{\mathcal A}_-^2&{\mathcal A}_+^2
\end{array}\right)
\left(\begin{array}{ccc}
{\rm e}^{+2r}\\
{\rm e}^{-2r}
\end{array}\right) \nonumber\\
&+&
\left(\begin{array}{ccc}
1-\left({\mathcal A}_+^2 + {\mathcal A}_-^2\right)\\
1-\left({\mathcal A}_+^2 + {\mathcal A}_-^2\right)
\end{array}\right).
\end{eqnarray}

\section{EIT filters}
\label{sect:eit}

In this section, we propose three kinds of EIT filters: (i)
low-pass, band-pass, and high-pass squeeze amplitude attenuation
filters, (ii) an S-shaped filter as a frequency-dependent squeeze
angle rotator, and (iii) an intra-cavity EIT filter as a
frequency-dependent squeeze angle rotator.

\begin{figure*}[t]
\begin{center}
\includegraphics[width=0.90\textwidth]{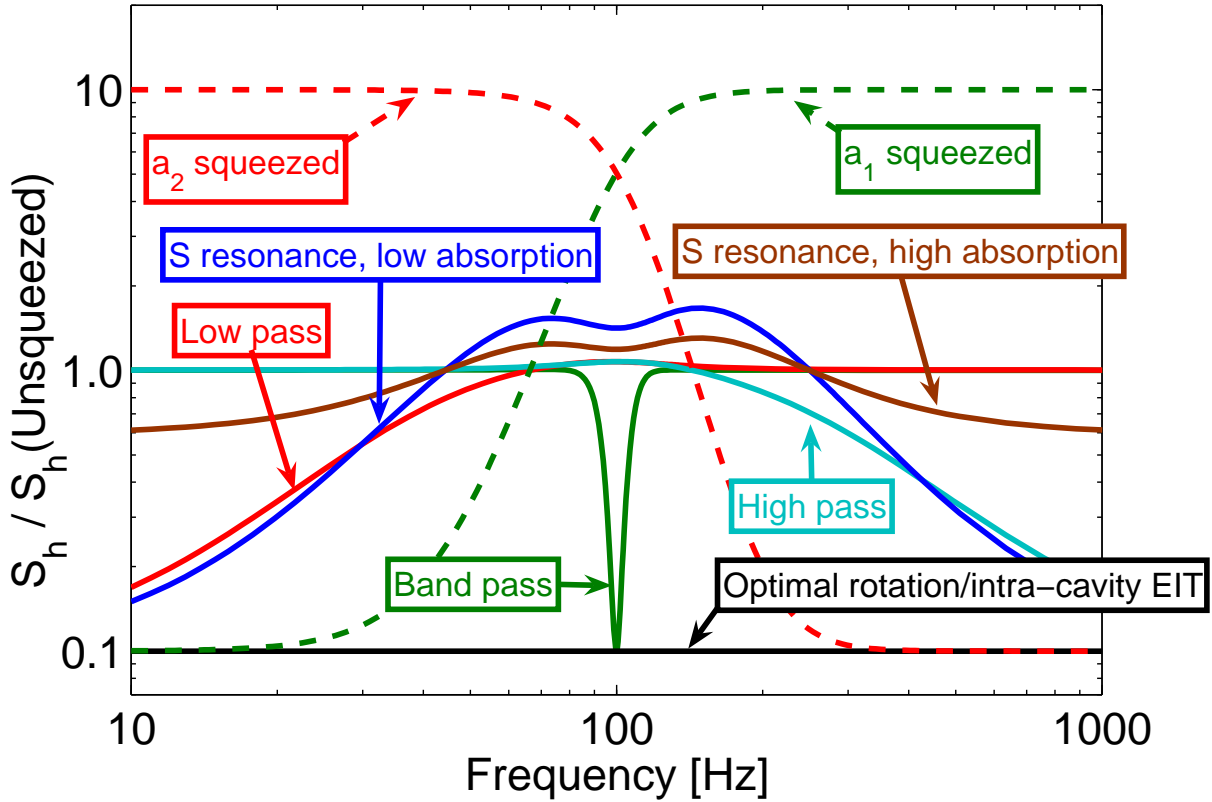}
\end{center}
\caption{The noise spectral density of a GW interferometer,
normalized to the noise density with no squeezed state injection.
We show the effect of low-pass (red), band-pass (green), high-pass
(cyan), S-shaped (blue for low and brown for high absorption
levels), and intra-cavity (black) filters and compare it to
frequency independent schemes. The dashed curves show unfiltered squeezing of the
$a_1$ (green) and $a_2$ (red) quadratures.
\label{fig:fig2} }
\end{figure*}

To obtain expressions for the EIT transmission ${\mathcal
T}(\pm\Omega)$, it is useful to use the following formulae for EIT
susceptibility, $\chi$~\cite{mikhailov04pra,taichenachev2003pra}
\begin{eqnarray}
\label{eq:chi1_eit}
\hspace{-1cm}\frac{\omega_0}{2\,c}\int_0^l\chi_1(\pm\Omega,z)\,
dz&=& A(\Delta)\,
\frac{\Gamma(\delta \pm \Omega)}{\Gamma^2+(\delta \pm \Omega)^2}
\nonumber\\
\hspace{0cm}&+&B(\Delta)\, \frac{\Gamma^2}{\Gamma^2+(\delta \pm \Omega)^2},\\
\label{eq:chi2_eit}
\hspace{-1cm}\frac{\omega_0}{2\,c}\int_0^l\chi_2(\pm\Omega,z) dz&=&
A(\Delta)\,
\frac{\Gamma^2}{\Gamma^2+(\delta \pm \Omega)^2} \nonumber\\
\hspace{0cm}&+& B(\Delta)\, \frac{\Gamma
(\delta \pm \Omega)}{\Gamma^2+(\delta \pm \Omega)^2} +C(\Delta),
\end{eqnarray}
where $\Gamma$ is the EIT resonance linewidth and the coefficients
$A$, $B$ and $C$ depend on the intensity and detuning, $\Delta$,
of the drive field. The first term on the right hand side in
Eq.~\eqref{eq:chi1_eit} corresponds to asymmetric phase dependence
on the two-photon detuning $\delta \pm \Omega$ and the second term
is responsible for symmetrical Lorentzian dependence. Similarly,
the first term in Eq.~\eqref{eq:chi2_eit} corresponds to
symmetrical dependence on the absorption coefficient of the EIT
medium, the second term indicates asymmetric absorption, and the
third term corresponds to broadband absorption in the medium.
Inserting Eqs.~\eqref{eq:chi1_eit} and \eqref{eq:chi2_eit} into
Eqs.~\eqref{eq:trans_chi1} and  \eqref{eq:theta_chi2}, we get
explicit expressions for ${\mathcal T}_{\pm}$ and $\theta_{\pm}$,
which are used calculate the noise at the output of the GW
interferometer $S_h$, as described in Section~\ref{sect:GWI}.

\subsection{Squeeze amplitude attenuation filters}

If we take $A=-C$, $B=0$, $C \geq 10$, and $\delta=0$
\cite{harris'97pt}, then
$\frac{\omega_0}{2c}\int_0^l\chi_2(\pm\Omega,z) dz \geq 10$ for
$\Omega \gg \Gamma$ and
$\frac{\omega_0}{2c}\int_0^l\chi_2(\pm\Omega,z) dz\simeq 0$ for
$\Omega \ll \Gamma$, and we obtain a symmetric EIT transmission
line with $A_{-} = 0$ and $\varphi_+ = 0$. This can be used as a
squeeze amplitude filter that retains squeezing at low
frequencies, but causes ordinary (unsqueezed) vacuum to replace
the anti-squeezed noise at higher frequencies that are outside the
EIT bandwidth. This is equivalent to the low-pass filters proposed
in~\cite{corbitt2004prd}. Similarly, buffer gas induced
electromagnetically induced absorption
(BGIEIA)~\cite{mikhailov04pra} may be used as a high-pass filter.
BGIEIA is similar to EIT, with a narrowband {\it absorption}
resonance instead of the transmission resonance of EIT. High-pass
filtering with BGIEIA can be realized with parameters $A \geq 10$,
$B=0$, $C=0$, and $\delta = 0$, such that
$\frac{\omega_0}{2c}\int_0^l\chi_2(\pm\Omega,z) dz \simeq 0$ for
$\Omega \gg \Gamma$ and
$\frac{\omega_0}{2c}\int_0^l\chi_2(\pm\Omega,z) dz \geq 10$ for
$\Omega \ll \Gamma$. A combination of two EIT resonances, equally
detuned from the carrier (obtained by Zeeman
splitting~\cite{wynands'99}), serves as a band-pass filter. In
Figure~\ref{fig:fig2} the effect of low-pass, band-pass, and
high-pass EIT filters on the noise spectral density of a GW
interferometer are shown by the curves labelled ``low pass''
(red), ``band pass'' (dark green), and ``high pass'' (cyan),
respectively. The noise spectra are normalized by the noise
spectral density of a conventional interferometer with no squeezed
state injection, which corresponds to the unity in
Figure~\ref{fig:fig2}. In each case, the harmful effects of
squeezing with a constant squeeze angle are reduced.

\subsection{Squeeze angle rotation filters}
The filters discussed so far only provide attenuation of the
squeezing, they do not produce any squeeze angle rotation. We now
introduce two cases of EIT that also produce squeeze angle
rotation.

\subsubsection{S-shaped EIT filter}

In principle, optimum squeeze angle rotation, corresponding to
$\theta \simeq -\Phi$, can be obtained with an asymmetrical -- or
S-shaped -- EIT filter, realized under conditions similar to
BGIEIA with parameters $A \simeq 0$, $B=-\pi/2$, and $C>|B|$. In
this case, the noise spectral density can be optimized over most
frequencies. However, as is evident from the curves labelled ``S
resonance'' (brown and blue) in Figure~\ref{fig:fig2}, the
improvement is small because of the high off-resonance losses
associated with the coefficient $C$; optimization would require
setting $C=0$. The noise is decreased at both high and low
frequencies, but worsened in the middle of the band due to two
effect. First, the imbalance in absorption ($A_- \neq 0$) between
the upper and lower sidebands required to obtain the squeeze angle
rotation causes the quadratures to be mixed and the noise may be
higher than shot noise. Second, the imperfect squeeze angle
rotation, $\theta \simeq -\Phi$, couples in noise from the
anti-squeezed quadrature.

\subsubsection{Intracavity EIT}

Placing an EIT medium in an optical cavity narrows the cavity
linewidth~\cite{muller97pra,lukin1998ol}. We consider a symmetric
EIT transmission resonance, such as the one used for our low-pass
filter, with a large linewidth $\Gamma \simeq 10~\rm{kHz}$, such
that ${\mathcal T}_{\pm} \simeq 1$, $\varphi_{+} \simeq 0$, and
$\varphi_- \simeq \pm \frac{\omega_0 l}{2c}\Omega
\left.\frac{d\chi_1}{d\Omega}\right|_{\Omega = 0}$ in the
frequency band of interest (100Hz). This configuration may be
understood in terms of the group velocity of the light through the
EIT medium
\begin{equation}
v_g = \frac{c}{\omega_0}\left(\left.\frac{d\chi_1}{d\Omega}\right|_{0}\right)^{-1}.
\end{equation}
We may then express
\begin{eqnarray}
\varphi_- \simeq \frac{l}{2 v_g}\Omega = \frac{\Omega l_{e}}{c},
\end{eqnarray}
where $l_{e} = \frac{c}{2 v_g}l$ is the effective cavity length.
The EIT medium serves as an additional delay line inside the
cavity. A group velocity as low as 8~m/s in a 12 cm long Rb vapor
cell was demonstrated by Budker {\it et al.}~\cite{budker99},
giving an effective length on the order of $10^6~\rm{m}$. To use
intra-cavity EIT as a squeeze angle rotation filter, we must
detune the cavity (but not the EIT medium) from the carrier. For
this case, the rotation arises from the detuned cavity, similar to
the filters in Ref.~\cite{kimble2002prd}, and not from the EIT.
The EIT acts only to modify the resonant linewidth of the cavity
by increasing its effective length. To achieve the required
$100$~Hz linewidth for the filter, we use an EIT with a large
effective length, combined with a short cavity. The performance of
this cavity is essentially identical to that of an isolated cavity
with no EIT and length equal to $l_{e}$, and may achieve optimal
squeeze angle rotation with a much smaller optical loss than a
traditional cavity of the same linewidth with the same real (not
effective) length, giving improved GW interferometer sensitivity
at all frequencies. We note that ${\mathcal T}_\pm \simeq 1$ is a
reasonable approximation because of the narrowed linewidth of the
system (${\mathcal T}_i \,v_g/l \ll \Gamma$, where ${\mathcal
T}_i$ is the transmission of the cavity input mirror). Detailed
calculations show that the effective loss of the system may even
be reduced by placing the EIT in a shorter cavity for some
parameter choices. The broadband reduction of noise from such an
intra-cavity EIT filter is shown in the curve labelled
``intra-cavity EIT'' (black) in Figure~\ref{fig:fig2}, and
corresponds exactly to the much coveted {\it optimal
frequency-dependent squeezing} of Ref.~\cite{kimble2002prd}.

\section{Conclusions}
\label{sect:conclusions}

In summary, we have shown that frequency-dependent squeeze
amplitude attenuation and squeeze angle rotation by EIT can
improve the sensitivity of GW interferometers. We find that the
EIT filters perform the same functions as previous filter designs,
but may be easier to implement, due to lower optical losses,
compactness, and easy linewidth tunability. While suitable EIT
media for the 1064nm transition -- the wavelength of laser light
in present GW interferometers --  have not yet been identified, we
note that the ideal wavelength for future GW interferometers may
well be determined by availability of high-power lasers,
low-absorption optical materials, squeezed light sources, high
quantum efficiency photodetectors, and perhaps EIT filters.\\

\section{Acknowledgment}
We would like to thank our colleagues at the LIGO Laboratory,
especially C. Wipf. We are grateful for valuable comments from I.
Novikova at Harvard University, A. B. Matsko at Jet Propulsion
Laboratory, P. K. Lam and K. McKenzie at Australian National
University, and G. M\"{u}ller at University of Florida. We
gratefully acknowledge support from National Science Foundation
grants PHY-0107417 and PHY-0457264.




\begin{thebibliography}{15}
\expandafter\ifx\csname natexlab\endcsname\relax\def\natexlab#1{#1}\fi
\expandafter\ifx\csname bibnamefont\endcsname\relax
  \def\bibnamefont#1{#1}\fi
\expandafter\ifx\csname bibfnamefont\endcsname\relax
  \def\bibfnamefont#1{#1}\fi
\expandafter\ifx\csname citenamefont\endcsname\relax
  \def\citenamefont#1{#1}\fi
\expandafter\ifx\csname url\endcsname\relax
  \def\url#1{\texttt{#1}}\fi
\expandafter\ifx\csname urlprefix\endcsname\relax\def\urlprefix{URL }\fi
\providecommand{\bibinfo}[2]{#2}
\providecommand{\eprint}[2][]{\url{#2}}

\bibitem[{\citenamefont{Caves}(1981)}]{caves1981prd}
\bibinfo{author}{\bibfnamefont{C.~M.} \bibnamefont{Caves}},
  \bibinfo{journal}{Phys. Rev. D} \textbf{\bibinfo{volume}{23}},
  \bibinfo{pages}{1693} (\bibinfo{year}{1981}).

\bibitem[{\citenamefont{Kimble et~al.}(2002)\citenamefont{Kimble, Levin,
  Matsko, Thorne, and Vyatchanin}}]{kimble2002prd}
\bibinfo{author}{\bibfnamefont{H.~J.} \bibnamefont{Kimble}},
  \bibinfo{author}{\bibfnamefont{Y.}~\bibnamefont{Levin}},
  \bibinfo{author}{\bibfnamefont{A.~B.} \bibnamefont{Matsko}},
  \bibinfo{author}{\bibfnamefont{K.~S.} \bibnamefont{Thorne}},
  \bibnamefont{and} \bibinfo{author}{\bibfnamefont{S.~P.}
  \bibnamefont{Vyatchanin}}, \bibinfo{journal}{Phys. Rev. D}
  \textbf{\bibinfo{volume}{65}}, \bibinfo{pages}{022002}
  (\bibinfo{year}{2002}).

\bibitem[{\citenamefont{Harms et~al.}(2004)\citenamefont{Harms, Schnabel, and
  Danzmann}}]{harms2004prd}
\bibinfo{author}{\bibfnamefont{J.}~\bibnamefont{Harms}},
  \bibinfo{author}{\bibfnamefont{R.}~\bibnamefont{Schnabel}}, \bibnamefont{and}
  \bibinfo{author}{\bibfnamefont{K.}~\bibnamefont{Danzmann}},
  \bibinfo{journal}{Phys.~Rev.~D} \textbf{\bibinfo{volume}{70}},
  \bibinfo{pages}{102001} (\bibinfo{year}{2004}).

\bibitem[{\citenamefont{Corbitt et~al.}(2004)\citenamefont{Corbitt, Mavalvala,
  and Whitcomb}}]{corbitt2004prd}
\bibinfo{author}{\bibfnamefont{T.}~\bibnamefont{Corbitt}},
  \bibinfo{author}{\bibfnamefont{N.}~\bibnamefont{Mavalvala}},
  \bibnamefont{and} \bibinfo{author}{\bibfnamefont{S.}~\bibnamefont{Whitcomb}},
  \bibinfo{journal}{Phys. Rev. D} \textbf{\bibinfo{volume}{70}},
  \bibinfo{pages}{022002} (\bibinfo{year}{2004}).

\bibitem[{\citenamefont{Budker et~al.}(1999)\citenamefont{Budker, Kimball,
  Rochester, and Yashchuk}}]{budker99}
\bibinfo{author}{\bibfnamefont{D.}~\bibnamefont{Budker}},
  \bibinfo{author}{\bibfnamefont{D.~F.} \bibnamefont{Kimball}},
  \bibinfo{author}{\bibfnamefont{S.~M.} \bibnamefont{Rochester}},
  \bibnamefont{and} \bibinfo{author}{\bibfnamefont{V.~V.}
  \bibnamefont{Yashchuk}}, \bibinfo{journal}{Phys. Rev. Lett.}
  \textbf{\bibinfo{volume}{83}}, \bibinfo{pages}{1767} (\bibinfo{year}{1999}).

\bibitem[{\citenamefont{Lukin et~al.}(1997)\citenamefont{Lukin, Fleischhauer,
  Zibrov, Robinson, Velichansky, Hollberg, and Scully}}]{lukin97prl}
\bibinfo{author}{\bibfnamefont{M.~D.} \bibnamefont{Lukin}},
  \bibinfo{author}{\bibfnamefont{M.}~\bibnamefont{Fleischhauer}},
  \bibinfo{author}{\bibfnamefont{A.~S.} \bibnamefont{Zibrov}},
  \bibinfo{author}{\bibfnamefont{H.~G.} \bibnamefont{Robinson}},
  \bibinfo{author}{\bibfnamefont{V.~L.} \bibnamefont{Velichansky}},
  \bibinfo{author}{\bibfnamefont{L.}~\bibnamefont{Hollberg}}, \bibnamefont{and}
  \bibinfo{author}{\bibfnamefont{M.~O.} \bibnamefont{Scully}},
  \bibinfo{journal}{Phys. Rev. Lett.} \textbf{\bibinfo{volume}{79}},
  \bibinfo{pages}{2959} (\bibinfo{year}{1997}).

\bibitem[{\citenamefont{Harris}(1997)}]{harris'97pt}
\bibinfo{author}{\bibfnamefont{S.~E.} \bibnamefont{Harris}},
  \bibinfo{journal}{Phys.\ Today} \textbf{\bibinfo{volume}{50}},
  \bibinfo{pages}{36} (\bibinfo{year}{1997}).

\bibitem[{\citenamefont{Xia et~al.}(1999)\citenamefont{Xia, Merriam,
  Sharpe, Yin, and Harris}}]{xia1999pra}
\bibinfo{author}{\bibfnamefont{H.}~\bibnamefont{Xia}},
  \bibinfo{author}{\bibfnamefont{A. J.} \bibnamefont{Merriam}},
  \bibinfo{author}{\bibfnamefont{S. J.} \bibnamefont{Sharpe}},
  \bibinfo{author}{\bibfnamefont{G. Y.} \bibnamefont{Yin}},
  \bibnamefont{and} \bibinfo{author}{\bibfnamefont{S. E.}
  \bibnamefont{Harris}}, \bibinfo{journal}{Phys. Rev. A}
  \textbf{\bibinfo{volume}{59}}, \bibinfo{pages}{R3190} (\bibinfo{year}{1999}).

\bibitem[{\citenamefont{Braje et~al.}(2003)\citenamefont{Braje, Bali\'{c},
  Yin, and Harris}}]{braje2003pra}
\bibinfo{author}{\bibfnamefont{D. A.}~\bibnamefont{Braje}},
  \bibinfo{author}{\bibfnamefont{V.} \bibnamefont{Bali\'{c}}},
  \bibinfo{author}{\bibfnamefont{G. Y.} \bibnamefont{Yin}},
  \bibnamefont{and} \bibinfo{author}{\bibfnamefont{S. E.}
  \bibnamefont{Harris}}, \bibinfo{journal}{Phys. Rev. A}
  \textbf{\bibinfo{volume}{68}}, \bibinfo{pages}{041801(R)} (\bibinfo{year}{2003}).

\bibitem[{\citenamefont{Akamatsu et~al.}(2004)\citenamefont{Akamatsu, Akiba,
  and Kozuma}}]{akamatsu2004prl}
\bibinfo{author}{\bibfnamefont{D.}~\bibnamefont{Akamatsu}},
  \bibinfo{author}{\bibfnamefont{K.}~\bibnamefont{Akiba}}, \bibnamefont{and}
  \bibinfo{author}{\bibfnamefont{M.}~\bibnamefont{Kozuma}},
  \bibinfo{journal}{Phys. Rev. Lett.} \textbf{\bibinfo{volume}{92}},
  \bibinfo{pages}{203602} (\bibinfo{year}{2004}).

\bibitem[{\citenamefont{Collett and Gardiner}(1984)}]{collett1984pra}
\bibinfo{author}{\bibfnamefont{M.~J.} \bibnamefont{Collett}} \bibnamefont{and}
  \bibinfo{author}{\bibfnamefont{C.~W.} \bibnamefont{Gardiner}},
  \bibinfo{journal}{Phys. Rev. A} \textbf{\bibinfo{volume}{30}},
  \bibinfo{pages}{1386} (\bibinfo{year}{1984}).

\bibitem[{\citenamefont{Mikhailov et~al.}(2004)\citenamefont{Mikhailov,
  Novikova, Rostovtsev, and Welch}}]{mikhailov04pra}
\bibinfo{author}{\bibfnamefont{E.~E.} \bibnamefont{Mikhailov}},
  \bibinfo{author}{\bibfnamefont{I.}~\bibnamefont{Novikova}},
  \bibinfo{author}{\bibfnamefont{Y.~V.} \bibnamefont{Rostovtsev}},
  \bibnamefont{and} \bibinfo{author}{\bibfnamefont{G.~R.} \bibnamefont{Welch}},
  \bibinfo{journal}{Phys. Rev. A} \textbf{\bibinfo{volume}{70}},
  \bibinfo{pages}{033806} (\bibinfo{year}{2004}).

\bibitem[{\citenamefont{Taichenachev et~al.}(2003)\citenamefont{Taichenachev,
  Yudin, Wynands, Stahler, Kitching, and Hollberg}}]{taichenachev2003pra}
\bibinfo{author}{\bibfnamefont{A.~V.} \bibnamefont{Taichenachev}},
  \bibinfo{author}{\bibfnamefont{V.~I.} \bibnamefont{Yudin}},
  \bibinfo{author}{\bibfnamefont{R.}~\bibnamefont{Wynands}},
  \bibinfo{author}{\bibfnamefont{M.}~\bibnamefont{Stahler}},
  \bibinfo{author}{\bibfnamefont{J.}~\bibnamefont{Kitching}}, \bibnamefont{and}
  \bibinfo{author}{\bibfnamefont{L.}~\bibnamefont{Hollberg}},
  \bibinfo{journal}{Phys. Rev. A} \textbf{\bibinfo{volume}{67}},
  \bibinfo{pages}{033810} (\bibinfo{year}{2003}).

\bibitem[{\citenamefont{Wynands and Nagel}(1999)}]{wynands'99}
\bibinfo{author}{\bibfnamefont{R.}~\bibnamefont{Wynands}} \bibnamefont{and}
  \bibinfo{author}{\bibfnamefont{A.}~\bibnamefont{Nagel}},
  \bibinfo{journal}{Appl.\ Phys.\ B} \textbf{\bibinfo{volume}{68}},
  \bibinfo{pages}{1 } (\bibinfo{year}{1999}).

\bibitem[{\citenamefont{M\"{u}ller et~al.}(1997)\citenamefont{M\"{u}ller,
  M\"{u}ller, Wicht, Rinkleff, and Danzmann}}]{muller97pra}
\bibinfo{author}{\bibfnamefont{G.}~\bibnamefont{M\"{u}ller}},
  \bibinfo{author}{\bibfnamefont{M.}~\bibnamefont{M\"{u}ller}},
  \bibinfo{author}{\bibfnamefont{A.}~\bibnamefont{Wicht}},
  \bibinfo{author}{\bibfnamefont{R.-H.} \bibnamefont{Rinkleff}},
  \bibnamefont{and} \bibinfo{author}{\bibfnamefont{K.}~\bibnamefont{Danzmann}},
  \bibinfo{journal}{Phys. Rev. A} \textbf{\bibinfo{volume}{56}},
  \bibinfo{pages}{2385 } (\bibinfo{year}{1997}).

\bibitem[{\citenamefont{Lukin et~al.}(1998)\citenamefont{Lukin, Fleischhauer,
  Scully, and Velichansky}}]{lukin1998ol}
\bibinfo{author}{\bibfnamefont{M.}~\bibnamefont{Lukin}},
  \bibinfo{author}{\bibfnamefont{M.}~\bibnamefont{Fleischhauer}},
  \bibinfo{author}{\bibfnamefont{M.}~\bibnamefont{Scully}}, \bibnamefont{and}
  \bibinfo{author}{\bibfnamefont{V.~L.} \bibnamefont{Velichansky}},
  \bibinfo{journal}{Opt. Lett.} \textbf{\bibinfo{volume}{23}},
  \bibinfo{pages}{295} (\bibinfo{year}{1998}).

\end{thebibliography}
\end{document}